\documentclass[11pt,twoside]{article}
\usepackage{asp2010}

\resetcounters

\begin{document}

\title{Juvenile Ultracool Dwarfs }
\author{Emily L.\@~Rice$^1$, Jacqueline K.\@~Faherty$^{1,2}$, Kelle Cruz$^{1,3}$, Travis Barman$^{4}$, Dagny Looper$^{5}$, Lison Malo$^{6}$, Eric E.\@~Mamajek$^{7}$, Stanimir Metchev$^{2}$, Evgenya L.\@~Shkolnik$^{8}$}
\affil{$^1$Department of Astrophysics, American Museum of Natural History, 79th~Street and Central Park West, New York, NY 10024, USA}
\affil{$^2$Department of Physics \& Astronomy, State University of New York, Stony Brook, Stony Brook, NY 11794, USA}
\affil{$^3$Department of Physics \& Astronomy, Hunter College, 695 Park Avenue, New York, NY 10065, USA}
\affil{$^4$Lowell Observatory, 1400 West Mars Hill Road, Flagstaff, AZ 86001, USA}
\affil{$^5$Institute for Astronomy, University of Hawaii at Manoa, 2680 Woodlawn Drive, Honolulu, Hawaii 96822, USA}
\affil{$^6$D{\'e}partement de Physique and Observatoire du Mont-M{\'e}gantic, Universit{\'e} de Montr{\'e}al (Qu{\'e}bec), H3C 3J7, Canada}
\affil{$^7$Department of Physics \& Astronomy, University of Rochester, 500 Wilson Boulevard, Rochester, NY 14627, USA}
\affil{$^8$Department of Terrestrial Magnetism, Carnegie Institute of Washington, 5241 Broad Branch Road, NW, Washington, DC 20015, USA}

\begin{abstract}
Juvenile ultracool dwarfs are late spectral type objects (later than $\sim$M6) with ages between 10~Myr and several 100 Myr. Their age-related properties lie intermediate between very low mass objects in nearby star-forming regions (ages 1--5 Myr) and field stars and brown dwarfs that are members of the disk population (ages 1--5~Gyr). Kinematic associations of nearby young stars with ages from $\sim$10--100~Myr provide sources for juvenile ultracool dwarfs. The lowest mass confirmed members of these groups are late-M dwarfs. Several apparently young L dwarfs and a few T dwarfs are known, but they have not been kinematically associated with any groups. Normalizing the field IMF to the high mass population of these groups suggests that more low mass (mainly late-M and possibly L dwarf) members have yet to be found. The lowest mass members of these groups, along with low mass companions to known young stars, provide benchmark objects with which spectroscopic age indicators for juvenile ultracool dwarfs can be calibrated and evaluated. In this proceeding, we summarize currently used methods for identifying juvenile ultracool dwarfs and discuss the appropriateness and reliability of the most commonly used age indicators. 
\end{abstract}

\section{Introduction}
Juvenile ultracool dwarfs are very low mass stellar and substellar objects. {\em Juvenile} refers to objects with intermediate ages (10~Myr~$<$~age~$\lesssim$~600~Myr). They lack substantial ongoing accretion and primordial circumstellar material, but they still exhibit some signatures of youth that are not seen in typical field objects. {\em Ultracool} dwarfs have a spectral type of $\sim$M6 or later. Juvenile ultracool dwarfs are typically identified by the combination of a late spectral type with one or more youth indicators: activity signatures, low gravity spectral features, membership in a young cluster or nearby moving group, and/or companionship to a known young star. A small but significant population of these objects are currently known, including such benchmark objects as 2MASS~J1207$-$39, a member of the $\sim$10~Myr~TW~Hydrae moving group and the host of a planetary-mass companion.

The properties of juvenile ultracool dwarfs play a role in many aspects of star formation and stellar evolution. A complete census of the low mass population of young, nearby moving groups is essential for understanding how the initial mass function varies across stellar environments. Characterization of the physical and circumstellar properties of juvenile ultracool dwarfs is crucial for complete understanding of any evolutionary phenomenon with a mass or age dependence, for example: planet formation, disk dissipation, angular momentum evolution, companion frequency, and chromospheric activity. Benchmark juvenile ultracool dwarfs (i.e. objects with well-characterized kinematic and physical properties) will essentially provide calibration data for evolutionary models. Finally, juvenile ultracool dwarfs provide excellent targets for exoplanet searches because they are nearby and young, thus potentially hosting self-luminous giant planets that provide a favorable contrast ratio and angular separation for direct imaging instruments \citep{Beichman10,Kataria10}.

The specific questions about juvenile ultracool dwarfs addressed in the splinter session were:

\begin{enumerate}
\item What is the most efficient and accurate method for identifying juvenile ultracool dwarfs and associating them with young nearby moving groups?
\item What properties/features are reliable age indicators for late-M, L, and T spectral types? 
\item How do juvenile ultracool dwarfs fit in with our current understanding of star formation, e.g., mass function, number density, multiplicity, disk fraction, etc.?
\end{enumerate}

The first question is addressed in Section~2 and the second in Section~3. Question 3 is not explicitly discussed in this proceeding. As a result of the splinter session it became clear that a more complete answer to 1 and 2 will further our understanding of point 3. Section~4 discusses important caveats for identifying young moving groups, evaluating membership, and using membership as an age indicator.

\section{Finding Low Mass Members}
\begin{figure}[!ht]
\plotone{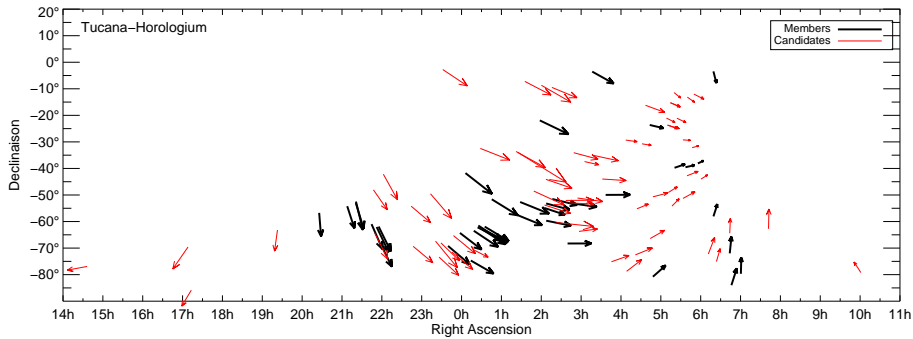}
\caption{\label{TucHorPM} Position on the sky and vector of proper motion for the known Tucana-Horologium members (black arrows) listed in \citet{Zuckerman04} and new candidates (red arrows) from Malo et al.\@~(in prep.). The size of the arrows is proportional to the proper motion amplitude. While distributed over a large fraction of the celestial sphere, all members follow a coherent and distinctive movement. A color version of this figure is available in the online edition.}
\end{figure}

The concept of moving groups emerged in the late 19th century when \citet{Proctor1869} and \citet{Huggins1871} noted that five of the A stars in the Ursa Major constellation were moving toward a common convergence point. Since that time, kinematic and activity-based studies have uncovered several other co-moving associations (e.g., Figure~\ref{TucHorPM}, \citealt{Eggen65,Eggen58}, \citealt{Zuckerman04}). The most studied of these to date include TW Hydrae, Tucana-Horologium, $\beta$~Pictoris, AB~Doradus, and $\eta$~Chamaeleontis, which are all nearby ($\lesssim$100 pc) and span ages from $\sim$10--100 Myr (see \citealt{Zuckerman04}, \citealt{Torres08}). Moving groups are older and more dispersed than star-forming regions with members widely spread-out on the sky. However, their proximity also makes them convenient laboratories for studying juvenile ultracool dwarfs because more distant ultracool dwarfs are too faint for detailed observations. 

\begin{figure}[!ht]
\plotone[width=0.9\hsize]{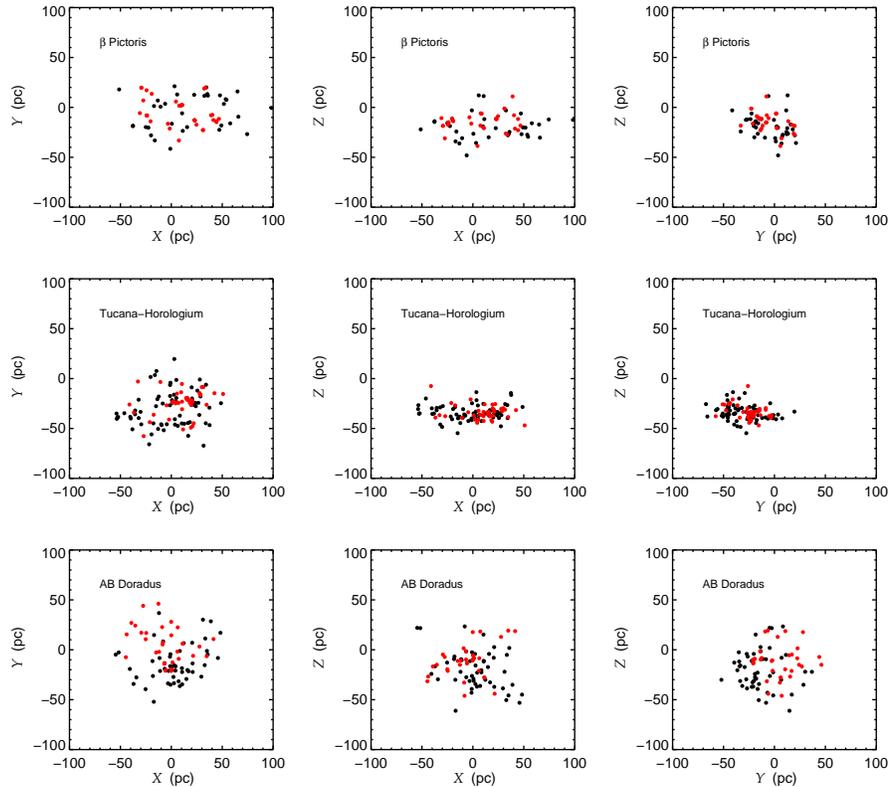}
\caption{\label{XYZ} Galactic position ($XYZ$) of known members (black) and new candidates (red) for three nearby young moving groups (Malo et al.\@~in prep.). A color version of this figure is available in the online edition.}
\end{figure}

Observational studies to discover new low-mass members are motivated by the apparent lack of M dwarfs in moving groups relative to the field initial mass function (\citealt{Torres06, Torres08}). However identifying and confirming low mass members can be difficult as these associations are sparse and widely dispersed on the sky. Age-indicative characteristics such as strong X-ray and H$\alpha$ activity, lithium absorption, and low surface gravity have been used as criteria for establishing youth among field objects (e.g. \citealt{McGovern04}, \citealt{Kirkpatrick08}, \citealt{West08}, \citealt{Cruz09}). Once proper motion, radial velocity, and distance are known, complete $UVW$ space motion and $XYZ$ positions can be used to robustly establish membership (e.g.,~Figure~\ref{XYZ}). However, parallaxes are time-consuming measurements rarely available for ultracool field objects; therefore, kinematic membership is often established without independent distance measurements. The high-resolution spectroscopy required to measure radial velocities and unambiguous youth indicators (H$\alpha$~10\% width, lithium absorption, alkali line equivalent widths, etc., see Section~3.1) are also time consuming for ultracool dwarfs. Caveats about evaluating the membership of objects with incomplete kinematic and spectral characterization are discussed in Section~4. 

\begin{figure}[!ht]
\plotone[width=0.75\hsize]{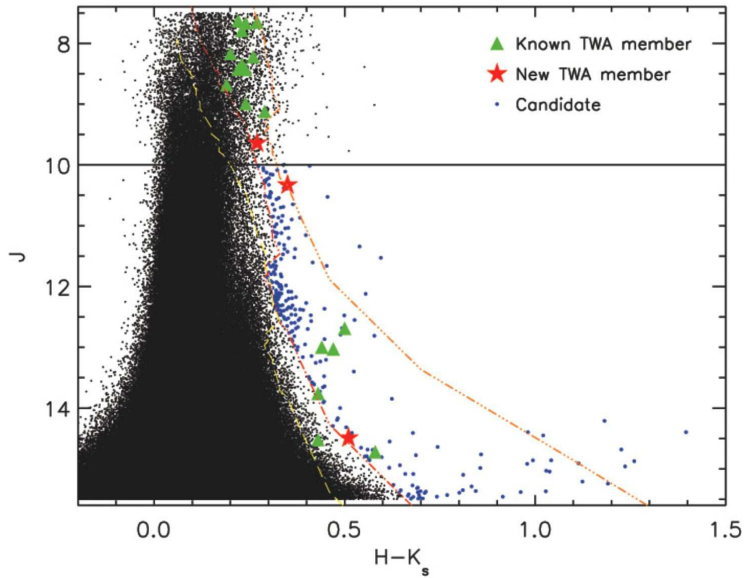}
\caption{\label{CMD} Near-infrared color-magnitude diagram for known (green triangles), new (red stars), and candidate (blue filled circles) TW~Hydrae members with isochronal tracks of \citet{Baraffe98} combined with \citet{Chabrier00} of 10~Myr at 10~pc (rightmost/orange dash-dotted line), 60~pc (middle/red dash-dotted line) and 100~pc (leftmost/yellow dashed line). Small black dots show the $>$800,000 targets after spatial and $J$-band magnitude selection. Figure from Looper et al.\@~2011, in preparation. A color version of this figure is available in the online edition.}
\end{figure}

Proper motion is available through numerous astrometric catalogs (e.g. USNO, LSPM-N, Hipparcos, Tycho, UCAC, etc.), but radial velocity measurements require high-resolution spectroscopy. Therefore, a number of studies have combined proper motion with near-IR or optical colors to search color-magnitude diagrams for new low mass members (e.g., \citealt{Montes01}, \citealt{Gizis02}, \citealt{Ribas03}, \citealt{Bannister2005}, \citealt{Clarke2010}). Follow-up spectroscopic observations to measure radial velocities and confirm kinematic association are only performed for high probability candidate members. In this manner, \citet{Lepine09} and \citet{Schlieder10} identify new members of $\beta$~Pictoris and AB~Doradus, \citet{RFC10} identify the lowest mass free-floating member of $\beta$~Pictoris, and Malo~et~al.\@~(in prep.) identify candidate members in Tucana-Horologium and employ a Bayesian model to evaluate the membership probability and the most probable distance based on measured properties (Figures~\ref{TucHorPM} and \ref{XYZ}). TW~Hydrae, with an age estimate as young as 8~Myr, is on the younger end of ``juvenile'', and some members still show evidence of accretion, although there is no associated molecular cloud \citep{Tachihara09}, and some members have debris disks. Looper et al.\@~(in prep.) identify new low mass members of the TW~Hydrae association (Figure~\ref{CMD}) including TWA~30A and~B, a low-mass co-moving system exhibiting signatures of an accretion disk and jet (\citealt{Looper10A,Looper10B}).

Juvenile ultracool dwarfs that are confirmed members of young groups are particularly important because their age is constrained by higher mass stars and properties of the group as a whole. Therefore their observed activity and spectroscopic properties can be used to calibrate models and constrain the ages of individual objects that lack a kinematic association. 

%\begin{figure}[!ht]
%\plotone{Looper_Fig1a.eps}
%\caption{\label{CMD} Near-infrared color-magnitude diagram for known (trangles), new (stars), and candidate (filled circles) TW~Hydrae members with isochronal tracks of \citet{Baraffe98} combined with \citet{Chabrier00} of 10~Myr at 10~pc (rightmost/orange dash-dotted line), 60~pc (middle/red dash-dotted line) and 100~pc (leftmost/yellow dashed line). Small black dots show the $>$800,000 targets after spatial and $J$-band magnitude selection. Figure from Looper et al.\@~2011, in preparation. A color version of this figure is available in the online edition.}
%\end{figure}

\section{Evaluating Spectral Age Indicators}

\subsection{M dwarfs}

\begin{figure}[!ht]
\plotone[width=0.7\hsize]{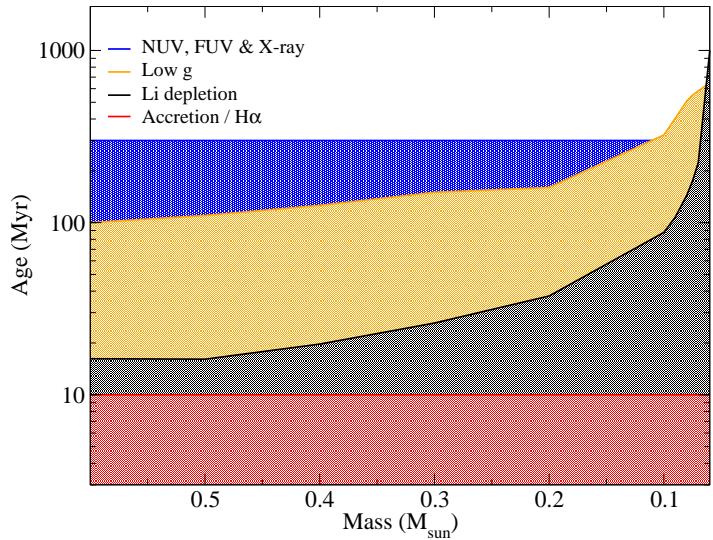}
\caption{\label{agediag} A summary of age diagnostics used in \citet{Shkolnik10}. Each technique provides an upper limit. The limits set by low gravity are from evolutionary models of \citet{Baraffe98} and lithium depletion from models of \cite{Chabrier96}. \cite{ByN03} set an upper limit of 10~Myr for a star still undergoing accretion. A color version of this figure is available in the online edition.
}
\end{figure}

There are several established age indicators for early-to mid-M dwarfs that can be applied to objects that are not (yet) kinematically associated with young moving groups. 
Figure~\ref{agediag} summarizes upper limits on age as a function of mass provided by four diagnostic properties: UV and X-ray emission, low surface gravity, lithium depletion, and accretion (as indicated by H$\alpha$ emission). 

X-ray and UV emission are related to magnetic activity, which can provide an upper age limit for early M dwarfs because magnetic activity is expected to decrease with age as angular momentum is dissipated over time \citep{Preibisch05}. However, for later spectral types ($\ge$M4) the activity lifetime is several Gyr, which is likely a consequence of the objects being fully convective and having a different mechanism for generating magnetic fields \citep{West08}. Nevertheless, activity evidenced by UV and/or X-ray emission has been successfully used to identify candidate members of nearby young moving groups \citep[e.g.,][]{Shkolnik09,Schlieder10}. The use of UV emission as an age diagnostic is less established than X-ray emission, but the sensitivity and sky coverage of the GALEX satellite compared to X-ray missions like {\em Chandra} and {\em ROSAT} enable promising early results \citep{Shkolnik10,Rodriguez10}. \citet{Shkolnik10} discovered two new mid-M~dwarf members of TW~Hydrae, TWA~31 and 32, using GALEX NUV and FUV emission.

\begin{figure}[!ht]
\plotone[width=0.7\hsize]{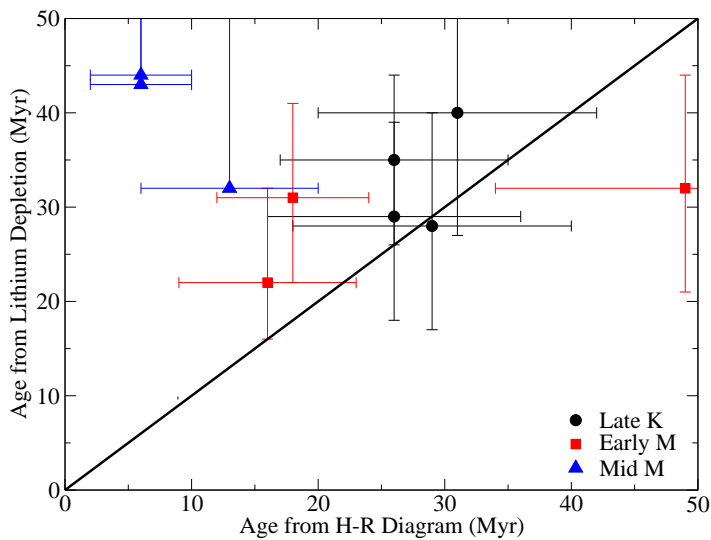}
\caption{\label{lithium} Data from \citet{Yee10} comparing the ages of 10 late-K to mid-M dwarfs derived from lithium depletion model versus those derived from H-R diagram \citep{Baraffe98}. For most objects the age from the lithium depletion models are substantially higher than the age inferred from the H-R diagram, lending support to the theory that increased frequency of episodic accretion can increase the efficiency of lithium burning, resulting in young objects having depleted their lithium earlier than would be expected from lithium depletion models that do not take episodic accretion history into account \citep{Baraffe10}. A color version of this figure is available in the online edition.
}
\end{figure}

Lithium abundance in low mass objects is a strong function of age, but the constraint it provides varies with mass. Lithium depletion models provide an age diagnostic that can be applied to individual objects via spectroscopic detection of lithium as well as to entire clusters via the determination of the lithium depletion boundary \citep[e.g.,][and references therein]{Mentuch08,Yee10}. A core temperature of 2.5~$\times$~10$^6$~K is required to burn lithium; therefore, objects with M~$<$~0.06~M$_{\odot}$ will never deplete their lithium. Thus for the lowest mass objects, lithium becomes a diagnostic of mass rather than age. Even for stars with M~$>$~0.06~M$_{\odot}$, the age determined by comparing measured lithium abundances to lithium depletion models is often inconsistent with the age determined from the H-R diagram (Figure~\ref{lithium}). A possible explanation for this discrepancy is found in \citet{Baraffe10}, who show that episodic accretion can temporarily increase core temperatures enough to burn lithium more efficiently. This results in prematurely depleted lithium compared to models that do not incorporate the effects of episodic accretion. However, the discrepancy between lithium age and H-R diagram age (Figure~\ref{lithium}) shows some mass dependence (later spectral types are typically more depleted in lithium than their H-R diagram ages imply), suggesting that there might still be a mass-dependent systematic uncertainty in the models (E.\@~Jensen, priv.\@~comm., 2010).

Several spectral features of ultracool dwarfs are gravity sensitive, including alkali lines (e.g., Na, K, e.g., \citealt{Gorlova03}), metal hydride bands (e.g., CaH, CrH, FeH, e.g., \citealt{Shkolnik09}), and metal oxide bands (e.g., VO, TiO, \citealt{Kirkpatrick08}). CaH in particular is used as a gravity indicator in M dwarfs, but weak CaH bands can be a result of high metallicity as well as low surface gravity. Many gravity-sensitive features are sensitive to temperature and/or metallicity so they must be interpreted with caution. Gravity-sensitive spectral features are discussed in more detail for L and T dwarfs below.

The strictest age constraint is obtained by detecting H$\alpha$ emission produced by ongoing accretion, providing an upper limit of 10~Myr \citep{ByN03}. H$\alpha$ emission can be reliably attributed to ongoing accretion (as opposed to chromospheric activity) if the width of the emission at 10\% the maximum strength is $\ge$~200 km~s$^{-1}$ \citep{White03}. Weak, narrow H$\alpha$ emission will persist for billions of years in most M~dwarfs as a result of chromospheric activity.

The spectroscopic and activity-based age indicators described above are very useful, but in and of themselves they are not failsafe methods of inferring the age of individual very low mass stars. The interpretation of many age indicators also depends on temperature, metallicity, and possibly more ambiguous properties like accretion history. Therefore it is necessary to approach age indicators with caution and to realistically assess the degeneracies and systematic uncertainties inherent in inferring the age of a very low mass star via spectral age indicators. 

\subsection{L and T dwarfs}

Age indicators are even more ambiguous and uncertain for L and T dwarfs, but important advances have been made in the past several years. Estimating the ages of substellar objects is complicated by their long cooling time and lack of a main sequence, which provides age-independent constraints on mass, luminosity, and effective temperature for hydrogen-burning stars. Brown dwarfs with M~$<$~0.06~M$_{\odot}$ will never reach temperatures high enough to burn lithium so the detection of lithium constrains mass instead of age for these objects. Furthermore, their cool, complex atmospheres include significant opacity from molecules and dust, inhomogeneous cloud structure, and non-equilibrium chemistry, further muddling the interpretation of their spectra and any potentially gravity-sensitive features.

\begin{figure}[!ht]
\plotone[width=0.75\hsize]{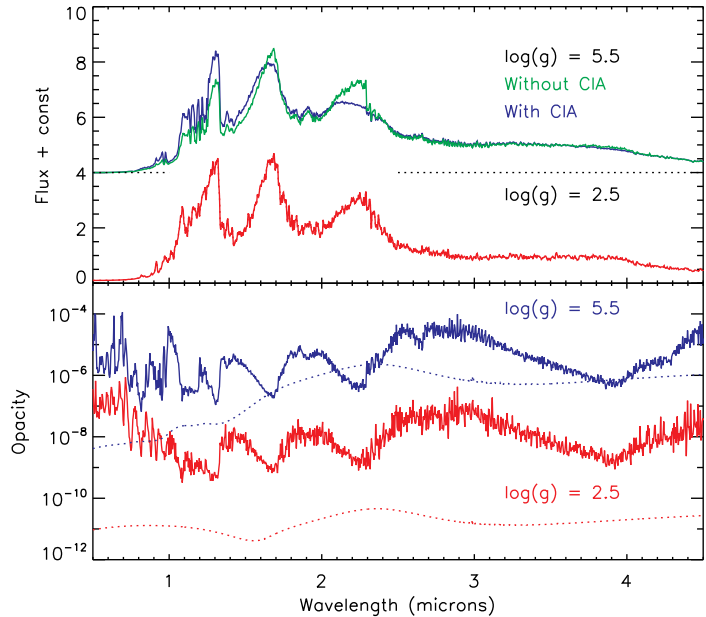}
\caption{\label{cia} Top: synthetic spectra calculated with the {\tt PHOENIX} model atmosphere code at two surface gravities, including and removing CIA from H$_2$ for the higher gravity. Bottom: opacity of H$_2$ CIA (dotted) and H$_2$O absorption (solid) at two surface gravities (Barman et al., in prep., after \citealt{Borysow97}). A color version of this figure is available in the online edition.}
\end{figure}

Substellar objects cool and shrink over their entire lifetimes, and gravity is the parameter that changes most with time \citep[e.g.,][]{Baraffe03}. Gravity and effective temperature uniquely determine age and mass, unlike luminosity which is degenerate with mass and age. There are several spectral features that are gravity-sensitive, but they are also typically dependent on temperature and/or metallicity, if not higher order parameters like dust, clouds, and chemistry. Nonetheless several gravity-sensitive spectral features are routinely used to identify young, low mass objects. The broadest feature is a peaked $H$-band spectral morphology, first observed by \citet{Lucas01} in spectra of substellar objects in Orion and later by \citet{Luhman05} for objects in IC~348 and by \citet{Allers07} for objects in Chamaeleon~II and Ophiuchus. %\citet{RFC10} note a strongly peaked $H$-band in the spectrum of 2MASS~0608$-$27, which they confirm as a member of the $\sim$12~Myr $\beta$~Pictoris moving group based on its $XYZ$ position and $UVW$ space velocity, indicating that the morphology is a useful diagnostic at intermediate ages. 
Figure~\ref{cia} shows the underlying physical explanation for the peaked $H$-band morphology. The feature is prominent for low surface gravity objects because the H$_2$O opacity dominates over collisionally-induced absorption (CIA) from H$_2$. For higher gravity objects, the opacity of the H$_2$ CIA is larger than the opacity from H$_2$O in the $H$- and $K$-bands, effectively flattening the peaks.

\begin{figure}[!ht]
\plotone[width=0.8\hsize]{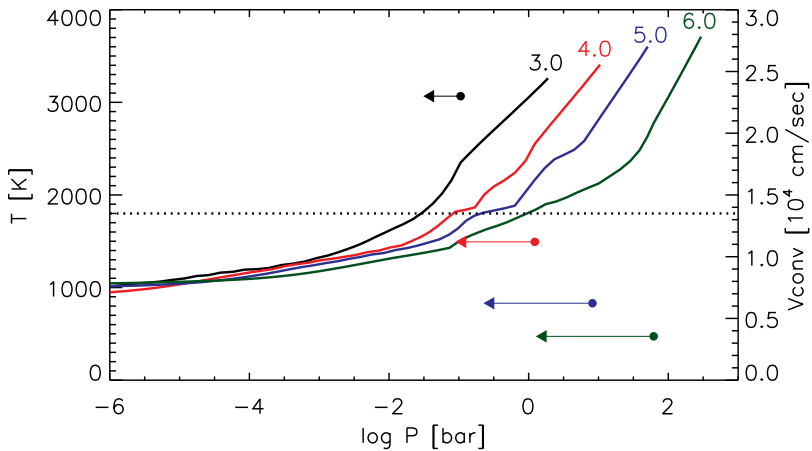}
\caption{\label{radconv} Temperature-pressure atmosphere profiles (solid lines, use left y-axis) for T$_{eff}$=1800~K and four values of log(g): 3 (black), 4 (red) , 5 (blue), 6 (green), calculated with the {\tt PHOENIX} code. The photosphere for each atmosphere is approximately where the solid line intersects the dotted line. The filled circles locate the radiative-convective boundary in pressure (x-axis), and the right vertical axis shows the maximum velocity just below this boundary in the convection zone. The length of the arrow shows the relative proximity of the radiative-convective boundary to the photosphere in pressure space. For lower surface gravities, the separation is smaller and the maximum convective velocity is higher. Thus, more efficient vertical mixing is expected at lower surface gravities, perhaps resulting in stronger non-equilibrium chemistry and thicker clouds (Barman et al., in prep.). A color version of this figure is available in the online edition.
}
\end{figure}

At moderate spectral resolutions ($R\gtrsim$1000), the strengths of alkali lines like Na {\sc i} and K {\sc i} have been shown to be sensitive to surface gravity, but they are also sensitive to temperature, resulting in a strong degeneracy that is evident at high resolution \citep{ZapOso04,Rice10}. Molecular features like CrH and VO have also been shown to be gravity-sensitive \citep{McGovern04,Kirkpatrick08,Cruz09}.

Two further properties of L and T dwarfs possibly related to youth are: red near-infrared colors and underluminosity. Red colors are expected to be linked to dust-enhanced atmospheres resulting from low surface gravity. Many unusually red (for their spectral type) objects found in the field show multiple signatures of youth \citep[e.g.,][]{Cruz09}. There are several objects with red colors lacking youth signatures \citep{Kirkpatrick10}, and high metallicity could also producer redder spectra \citep{Burrows06,Looper08}. While overluminosity on a color-magnitude diagram is a hallmark for youth in low mass stars \citep{Luhman07}, young L and T dwarfs appear {\em under}luminous \citep{Metchev06}. Moreover, from a parallax survey of eight low surface gravity L~dwarfs, Faherty et al.\@~(in prep.) determine that these objects are $\sim$1 magnitude underluminous on a brown dwarf near-IR H-R diagram.

Disentangling low gravity and other spectroscopic youth indicators becomes even more problematic at and beyond the L-T transition. It is becoming apparent that low gravity objects with effective temperatures comparable to known T~dwarfs (e.g., the young planetary-mass object 2MASS~J1207$-$39b) have L dwarf spectral types because they lack CH$_4$ absorption. However, the atmosphere is probably lacking CH$_4$ not because the atmosphere is too hot but because low gravity strengthens the effects of vertical mixing (Figure~\ref{radconv}). This issue is particularly important for wide, self-luminous extrasolar planets for which low resolution near-infrared spectra can now be obtained, like HR~8799b \citep{Bowler10}.

\section{Caveats for Young Groups and their Members}

Because of the difficulty in assigning ages to isolated field objects \citep{Mamajek09,Soderblom10}, young stellar groups play an important role in studies of age-dependent phenomena. Groups are observed (or assumed) to be approximately coeval, and group membership is usually used as a \emph{primary} indicator of age. However, membership must be very carefully evaluated when used as an age indicator. 

Assigning membership to an object and adopting the group age should be done cautiously and with as much corroborating evidence as possible. Other youth indicators such as rotation, activity, lithium, low gravity, full three-dimensional kinematics (radial velocity and parallax in addition to proper motion) and common proper motion with another member should also be considered.

\begin{figure}[!ht]
\plottwo{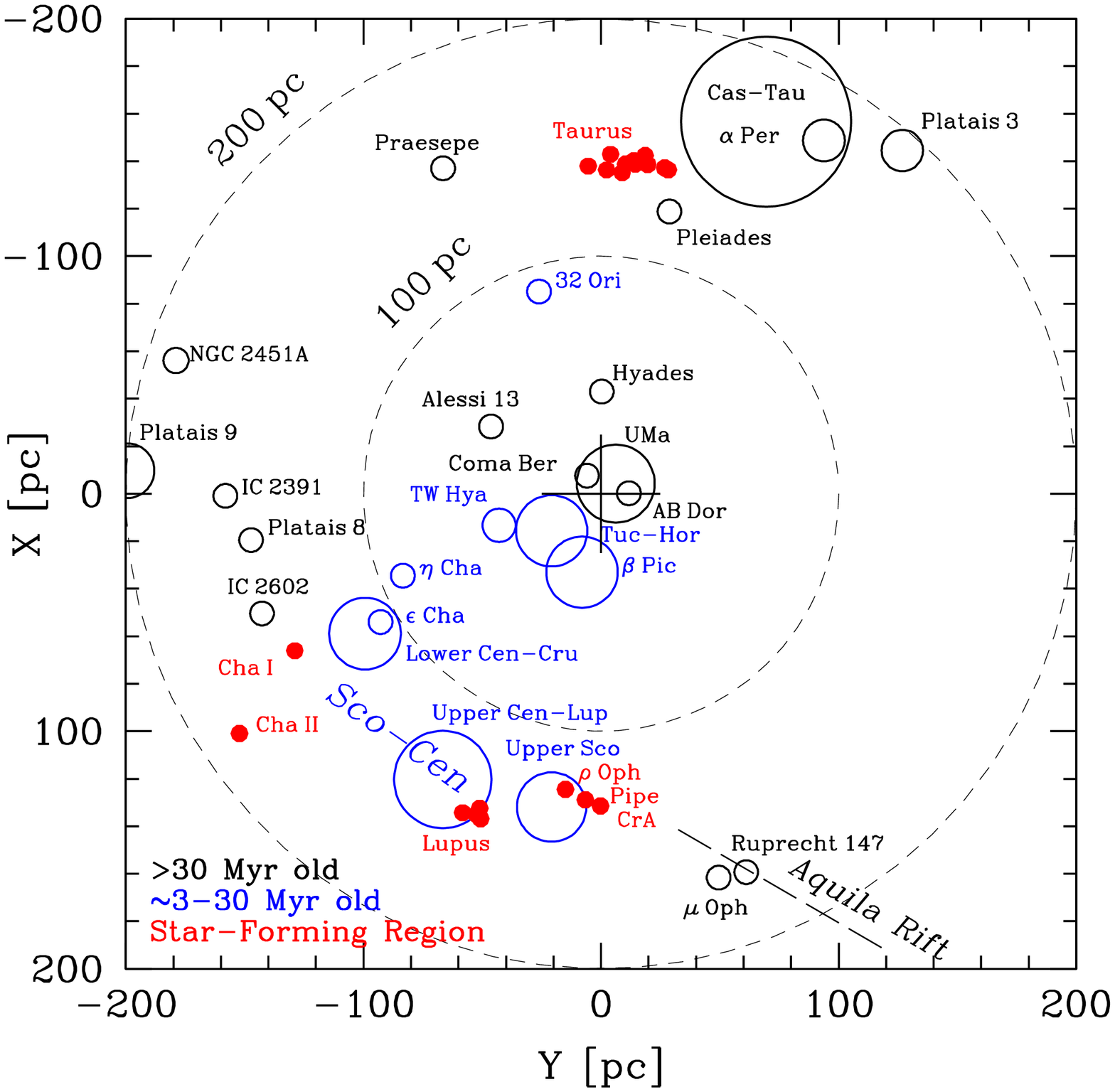}{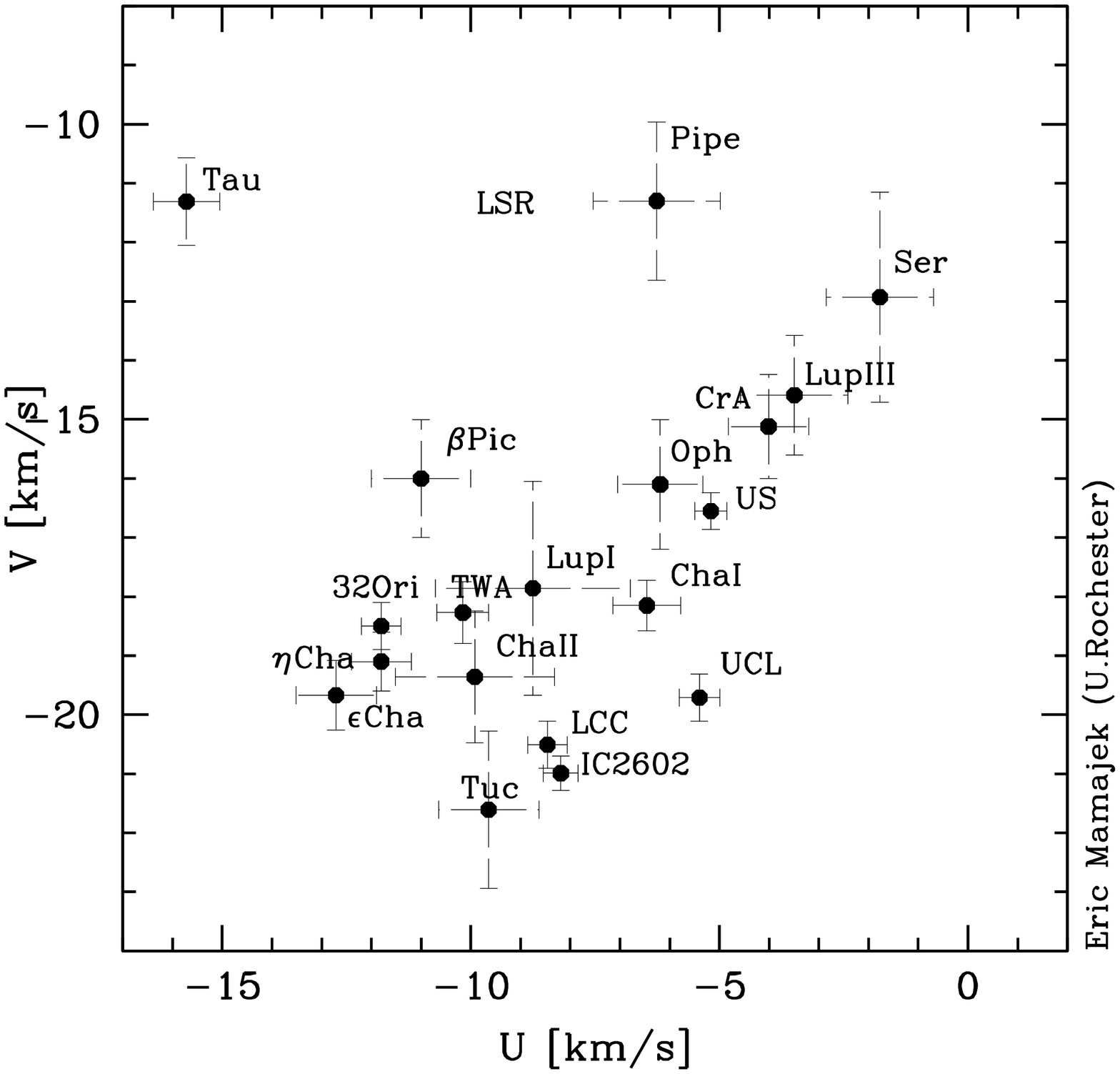}
\caption{\label{mama} Left: $XY$ distributions of star-forming regions and young groups within 200~pc of the Sun. Right: $UV$ distributions of star-forming regions and young groups. The close proximity of many groups in $XYZ$ and $UV$ requires that membership be evaluated carefully, particularly when kinematic and distance measurements are incomplete (Mamajek~2011, submitted). A color version of this figure is available in the online edition.}
\end{figure}

In particular, caution is urged when assigning membership with just proper motions because stellar groups of different ages can have similar velocities. As shown in Figure~\ref{mama}, the $UV$ distributions of the nearest, youngest groups are tightly clustered. However, the obvious nuclei of these groups all have velocity dispersions of only $\sim$1 km~s$^{-1}$, independent of density. In order to reliably assign membership, complete and accurate $UVW$ velocities and $XYZ$ coordinates are needed.

Furthermore, similar velocities are not sufficient to warrant the definition of a new group. Stars with consistent space motions could be a ``supercluster" kinematic stream and not related to formation at all. Superclusters are now known to be dynamical streams in the Milky Way galaxy as a result of spiral density waves, and the common motion of constituent stars does not imply a common age \citep[e.g.,][]{Famaey05}. Certain proposed young moving groups are unphysical -- that is, do not share a common or age -- because of large scatters in their H-R diagrams, radial velocities, distances, and/or peculiar velocities. An upcoming study of the revised Hipparcos astrometry for young stellar groups within 100~pc by Mamajek~2011 (submitted) shows that some candidate groups appear to be unphysical: Chereul~2, Chereul~3, Latyshev~2, and Polaris \citep{Chreul99,Latyshev77, Turner04}.

\acknowledgements The authors would like to thank the organizers of the Cool Stars 16 meeting, particularly Suzanne Hawley, the head of the SOC, for providing the opportunity to have this splinter session; Adam Burgasser, the SOC liaison, for helping us organize it; and the participants for engaging in a productive discussion.

\bibliography{Rice_E}

\end{document}